\documentclass[fleqn,10pt]{wlscirep}
\usepackage[utf8]{inputenc}
\usepackage[T1]{fontenc}

\usepackage{graphicx}
\usepackage{dcolumn}
\usepackage{bm}
\usepackage{textgreek}
\usepackage{stackengine}
\usepackage{braket}
\usepackage{xr}
\usepackage{cleveref}

\title{Efficient extreme-ultraviolet high-order wave mixing from laser-dressed silica}

\author[1,*]{Sylvianne D.C. Roscam Abbing}
\author[1]{Filippo Campi}
\author[1]{Brian de Keijzer}
\author[2]{Corentin Morice}
\author[1]{Zhuang-Yan Zhang}
\author[1]{Maarten L.S. van der Geest}
\author[1,3,*]{Peter M. Kraus}
\affil[1]{Advanced Research Center for Nanolithography, Science Park 106, 1098 XG Amsterdam, The Netherlands}
\affil[2]{Institute for Theoretical Physics Amsterdam and Delta Institute for Theoretical Physics, University of Amsterdam, 1098 XH Amsterdam, Netherlands}
\affil[3]{Department of Physics and Astronomy, and LaserLaB, Vrije Universiteit, De Boelelaan 1081, 1081 HV Amsterdam, The Netherlands}

\affil[*]{s.roscam@arcnl.nl; p.kraus@arcnl.nl}

\begin{abstract}
The emission of high-order harmonics from solids \cite{ghimire11a,schubert14a,luu15a,golde08a} under intense laser-pulse irradiation is revolutionizing our understanding of strong-field solid-light interactions \cite{ghimire11a,schubert14a,luu15a,vampa15b,yoshikawa17a,hafez18a,jurgens20a}, while simultaneously opening avenues towards novel, all-solid, coherent, short-wavelength table-top sources with tailored emission profiles and nanoscale light-field control\cite{franz19a,roscamCLEO21}.
To date, broadband spectra have been generated well into the extreme-ultraviolet (XUV) \cite{luu15a,luu18b,han19a,uzan20a}, but the comparatively low conversion efficiency still lags behind gas-based high-harmonic generation (HHG) sources \cite{luu15a,luu18b}, and have hindered wider-spread applications. 
Here, we overcome the low conversion efficiency by two-color wave mixing. A quantum theory reveals that our experiments follow a novel generation mechanism where the conventional interband and intraband nonlinear dynamics are boosted by Floquet-Bloch dressed states, that make solid HHG in the XUV more efficient by at least one order of magnitude. Emission intensity scalings that follow perturbative optical wave mixing, combined with the angular separation of the emitted frequencies, make our approach a decisive step for all-solid coherent XUV sources and for studying light-engineered materials.

\end{abstract}

\begin{document}

\flushbottom
\maketitle

\thispagestyle{empty}

\section*{Main Text}
High-harmonic generation from gases has revolutionized ultrafast spectroscopy \cite{kraus18a,kraus18b}, imaging science \cite{miao15a,witte14a}, and is finding its way into first industrial applications \cite{ku16a,roscam20a,roscam21a}, but is still limited by the low conversion efficiency from infrared to XUV photons. A thorough understanding of the mechanism of gas-based HHG spurred the use of sculpted multi-color drivers that favorably modify the ionization rate, and thus increase the conversion efficiency \cite{kim05a,brizuela13a,haessler14a,rajeev16a,roscam20a}.  
The mechanisms of solid HHG are still being debated. Different generation regimes exist, and are characterized by the choice of the driving field and generation material. Proposed mechanisms include nonlinear intraband currents \cite{ghimire11a,luu15a}, arising from carrier acceleration by the strong field after band-gap excitation, electron-hole recollisions after carrier excursion \cite{schubert14a,vampa15b} creating an interband polarization, ionization-induced injection currents \cite{jurgens20a} for low-order harmonics, 
or treating emission as originating from dressed eigenstates in a strong static field, which under high excitation intensities drives Bloch oscillations, known as Wannier-Stark states \cite{higuchi14a}.
The importance of carrier-injection dynamics in solid HHG \cite{jurgens20a}, combined with the observation of enhanced gas HHG for short-wavelength drivers, which favorably modifies the ionization step \cite{lai13a} and thus enhances the conversion efficiency, points towards a new strategy for both enhancing the efficiency of solid HHG, as well as shedding light on the generation mechanism: two-color non-collinear XUV generation.
\\

\begin{figure*}[t]
\includegraphics[width=\textwidth,keepaspectratio]{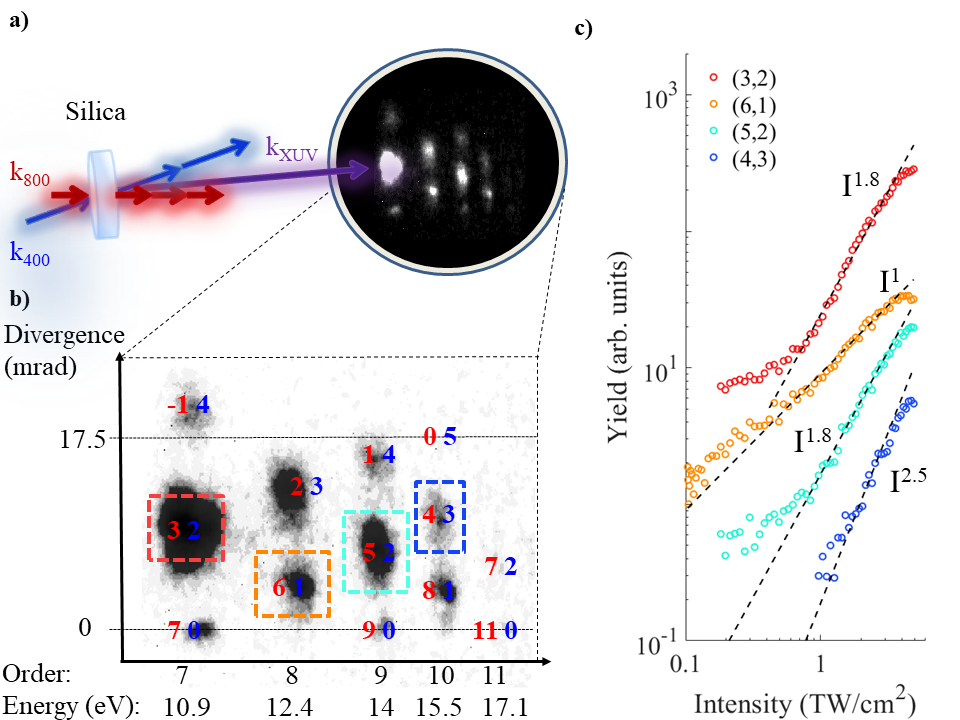}
\caption{High-order wavemixing in silica. a) Experimental arrangement. b) Far-field spectrum of XUV wave mixing in silica. WMOs are spectrally dispersed in the horizontal direction and propagate freely in the vertical direction. WMOs generated only from 800~nm or 400~nm propagate at 0~mrad or 17.5~mrad, respectively. Harmonic orders refer to multiples of the 800-nm fundamental. c) Yield of different WMOs as a function of the intensity of the 400-nm pulse. The dashed lines are power laws, whose exponent is indicated in the plot. The different lines are vertically offset for clarity. The intensity of the 800-nm pulses was 11 TW/cm\textsuperscript{2}. }
\label{fig:fig0}
\end{figure*}

Similar to non-collinear HHG in gases \cite{bertrand11a,heyl14a}, the angle between fundamental (800~nm) and second harmonic (SH) (400~nm) driving pulses (Fig.~\ref{fig:fig0}a) provides an angular separation of the photon channels and prevents their overlap and interference, unlike in collinear two-color HHG \cite{li18a}.
Non-collinear HHG leads to an emission pattern (Fig.~\ref{fig:fig0}b) of harmonic wave mixing orders (WMOs) labeled $(n,m)$, which follow momentum conservation and can thus be understood by wave vector addition of the underlying combination of $n$ fundamental and $m$ second-harmonic photons. Only combinations of an odd total number of photons are dipole-allowed, due to the centrosymmetry of amorphous silica. More details are given in the supplementary information (SI), section I.
The emission patterns in Figs.~\ref{fig:fig0}a,b show strongly enhanced WMOs compared to the angularly separated emission from the respective fundamental beams, which are indicated by dashed lines in Fig.~\ref{fig:fig0}b.
The separation of fundamental and SH photon channels in each WMO allows for identifying the effect of intensity scaling of the pulses on the harmonic emission, as shown in Fig.~\ref{fig:fig0}c for the scaling of the 400-nm pulse intensity - a first step in investigating and optimizing the photon-yield enhancement apparent in the WMOs. 
The yield of all WMOs faithfully follows a power-law trend over a large intensity range. The exponents of these power-law scalings (i.e. the slopes in a log-log plot) are close to the number $m$ of 400-nm photons involved, which would be the exponent for the scaling of the yield as a function of intensity in perturbative optical wave mixing. 

\begin{figure*}[t]
\includegraphics[width=\textwidth,keepaspectratio]{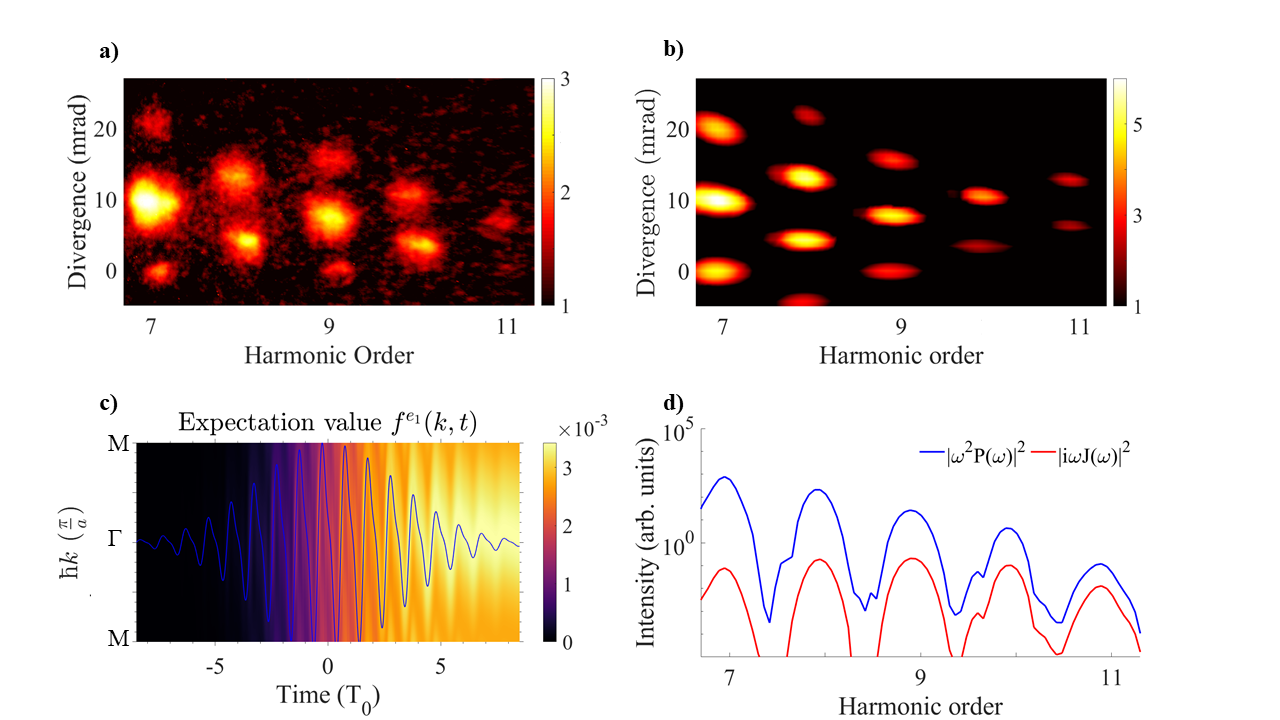}
\caption{Simulation of solid-state high-harmonic generation. a) Recorded far-field spectrum, the color axis represents the yield on a logarithmic scale. b) Simulated far-field spectrum, the color axis represents the yield on a logarithmic scale. c) Simulated electron population $f^{e_1}(k,t)$ (normalized magnitude given by color axis) of the first conduction band along the $\Gamma$-M crystal direction, in momentum and time space. The blue line represents the vector potential of the laser pulse. $T_0$ denotes the cycle duration of the 800-nm pulse (2.7 fs). d) Comparison between the contributions of the interband polarization (blue) and intraband current (red).} \label{fig:fig2}
\end{figure*}

In order to understand the origin of the enhanced WMOs (Fig.~\ref{fig:fig0}b) and the perturbative intensity scalings (Fig.~\ref{fig:fig0}c) we solve the semiconductor Bloch equations \cite{lindberg1988a} to simulate solid-state harmonic generation.
The field-driven carrier population exchange between the bands leads to the generation of an interband polarization and an intraband current (Supplementary Information [SI], sec.II A). This model was solved for a grid of points, which was spanned by the crossed 800-nm and 400-nm pulses that form a spatial interference grating. The resulting spatially modulated XUV emission was then propagated to the far-field by Fraunhofer diffraction (Fig.~\ref{fig:fig2}b), resulting in an excellent reproduction of the experimental wave mixing pattern (Fig.~\ref{fig:fig2}a).  
Zooming into the attosecond electron dynamics in k-space that drive HHG, we observe that the intense two-color laser field excites carriers from the valence band to the first conduction band not just at the $\Gamma$-point but all across $k$-space. Simultaneously, the field accelerates the carriers along the energy bands, resulting in carrier trajectories in $k$-space that trace out the vector potential of the field (Fig.~\ref{fig:fig2}c). Due to the high ponderomotive energy the carriers are accelerated beyond the first edge of the Brillouin zone and undergo Bloch oscillations. The concentration of carriers at a given $k$-point undergoes rapid changes in time, inducing a high-frequency interband polarization that radiates light at high photon energies, see Fig.~\ref{fig:fig2}d. The radiation due to the carrier acceleration inside the bands, the intraband current, is significantly lower than the radiation from interband polarization for the harmonic energies measured in this manuscript. 
Both the interband and intraband mechanisms have in common that they are induced by laser-driven motion of carriers inside the bands. To identify if this laser-driven carrier motion gives rise to the perturbative intensity scalings, we model the harmonic emission originating only from the electron group velocity, which we describe semi-classically \cite{wegener2005extreme,luu15a} using a time-dependent crystal momentum which traces out the shape of the band during acceleration (SI, sec.II E). 


\begin{figure*}[b!]
\includegraphics[width=\textwidth,keepaspectratio]{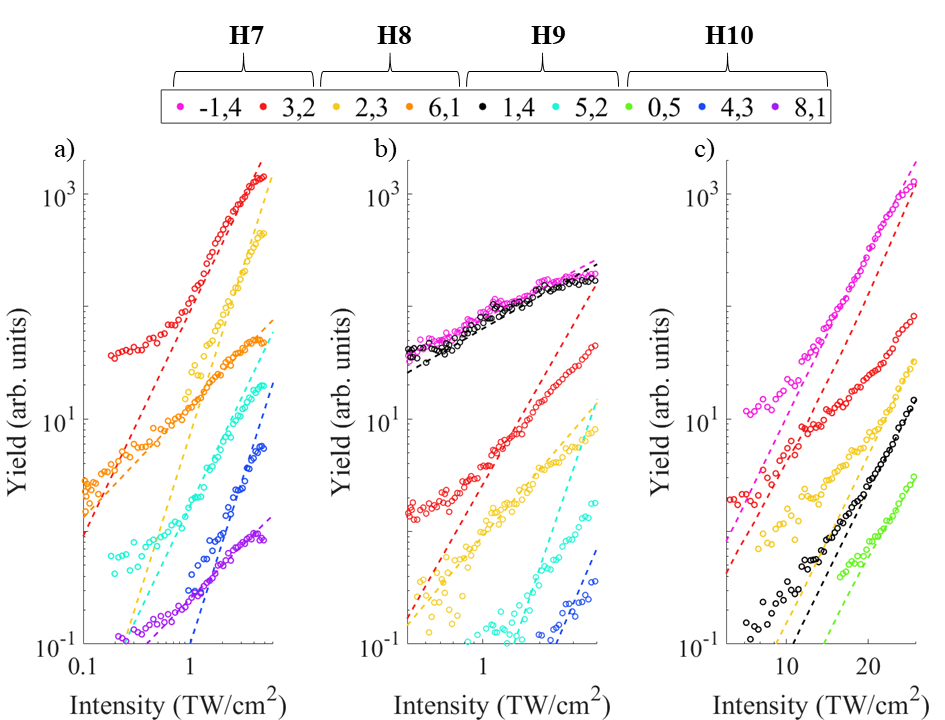}
\caption{Experimental (open circles) and simulated (dashed lines) intensity scaling of different harmonic orders. a) only the intensity of the 400-nm pulse is scaled (the ratio I\textsubscript{400}/I\textsubscript{800} varies between 0.01:1 to 0.4:1 from left to right on the horizontal intensity axis, I\textsubscript{800}=11.0 TW/cm\textsuperscript{2})  b) only the intensity of the 800-nm pulse is scaled (the ratio I\textsubscript{800}/I\textsubscript{400} varies between 0.06:1 to 0.7:1 from left to right on the horizontal intensity axis, I\textsubscript{400}=5.4 TW/cm\textsuperscript{2}), and c) the total intensity of two colors is scaled (the ratio I\textsubscript{400}/I\textsubscript{800} is 4.3:1). The color coding is consistent across the panels.}
\label{fig:fig3}
\end{figure*}

A detailed investigation of the experimental (open circles) and semi-classically simulated (dashed lines) yields of the WMOs as a function of intensity of the driving fields is shown in Figure.~\ref{fig:fig3}.
Figure \ref{fig:fig3}a shows a scaling of the 400-nm pulses, wherein the intensity of the fundamental 800-nm field is kept constant. The dashed lines in Fig. \ref{fig:fig3}a show that the simulated trend of a specific harmonic order $(n,m)$ reproduces the highly nonlinear dependence of the scaled driving field. We observe saturation of the XUV yield at high intensities, when approaching the damage threshold.
Next, experiments and simulations were performed for constant intensity of the 400-nm pulse, while the energy of the fundamental pulse is increased in steps. The behaviour in both experiment and simulations  resembles that of a perturbative process. At the higher intensities, the fundamental and the SH foci may have lost some overlap, which leads to an overestimation of the 800-nm intensity involved in the wave mixing process and therefore lowers the power dependence. WMOs that contain more 800-nm~photons deviate more severely from a pure power law, as observed for WMOs (4,3) and (5,2).
Figure \ref{fig:fig3}c shows a scaling of the total intensity, where the energy of the two pulses is increased independently in steps, to keep the ratio of intensities constant across the scan. The experimental data again show good agreement with semi-classical predictions, and reproduce the nonlinear dependence. More specifically, most orders follow a power law of 5 for these 5\textsuperscript{th} order processes shown in Fig.\ref{fig:fig3}c. The yield as a function of intensity is reproduced well for the entire intensity range. The scaling law of WMO (3,2) systematically shows a lower power dependence, possibly caused by a drift in the beams at higher intensities, which leads to an overestimation of the intensity. Intensity scalings of WMOs that contain fewer 800-nm photons are less affected by the drift. 
The good agreement between experiments and simulations in Fig.~\ref{fig:fig3} confirms that the laser-driven carrier motion drives the XUV generation of solid-state HHG, resulting in a perturbative scaling for large intensity ranges. This has important implications for XUV sources based on solid HHG. The perturbative response causes the yield of solid HHG as a function of driving intensity to rise with an exponent determined by the number of photons $(n,m)$ involved in HHG. This enables selective enhancement of individual or a selection of WMOs, depending on the relative strength of the two driving fields. Together with the natural angular separation of WMOs, this illustrates that XUV wave mixing in solids can help to realize an all-solid single-element XUV generator and multi-wavelength beamsplitter, where different energy emissions are mapped onto different emission angles. Further separation of the remaining energies in a beam could be realized by the simple addition of a hard aperture, without the need for lossy metal filters. 

\begin{figure*}[!]
\includegraphics[width=\textwidth,keepaspectratio]{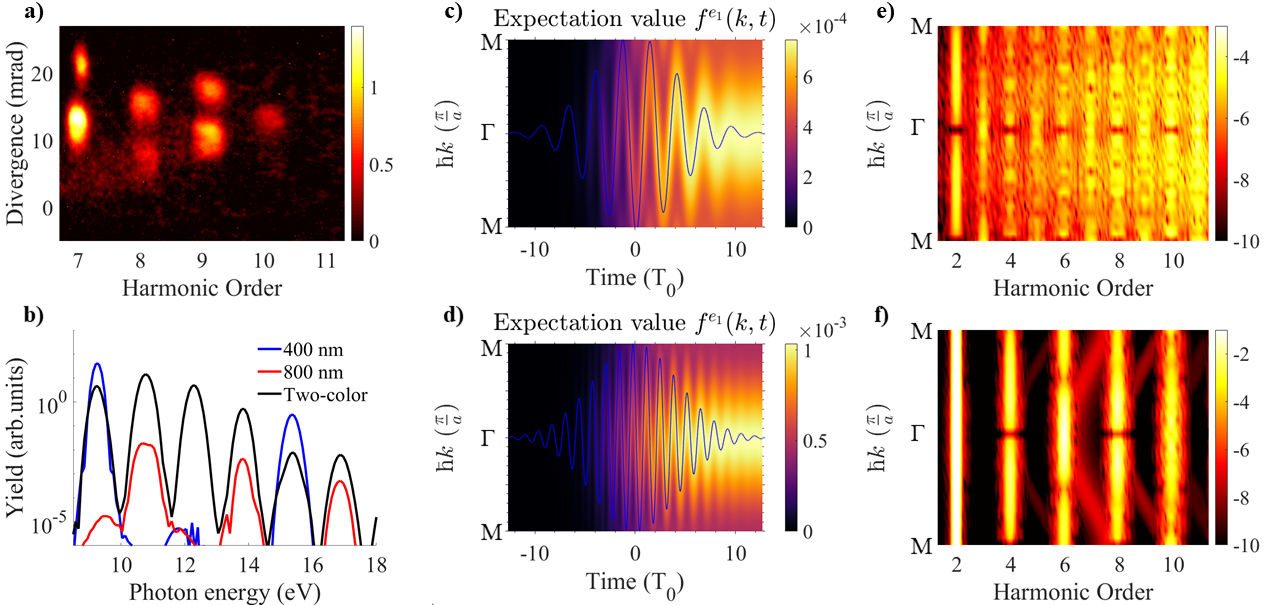}
\caption{Comparison of 800-nm 400-nm driven solid-state high-harmonic generation a) Wave mixing with equal 400-nm and 800-nm intensity of 5.4 TW/cm\textsuperscript{2}. b) Simulated spectra, comparing 400-nm, 800-nm and two-color driving with equal intensity. c) Electrons in the first conduction band as a function of time and crystal momentum for an 800-nm pulse (vector potential represented by the blue line), and d) for a 400-nm pulse. e) Energy- and momentum-resolved polarization for 800~nm, and f) for 400~nm, both for an intensity of 1.6 TW/cm\textsuperscript{2}. The color bar represents harmonic yield on a logarithmic scale. Harmonic orders refer to multiples of the 800-nm fundamental, even orders of the driving frequencies cancel out in the emission spectrum due to symmetry.} \label{fig:fig4}
\end{figure*}

Additionally, we deduce from experimental wave-mixing patterns a dramatic yield enhancement for WMOs, as compared to single-color harmonics. In Fig. \ref{fig:fig2}a, WMO (3,2) is the most intense emission, whereas (5,2) is the most intense at a photon energy of 14 eV, and (7,2) is the most intense at 17.1 eV. Comparing the increase in yield over increasing driving intensity between a certain WMO and the on-axis harmonic at the same photon energy (SI, section III.C), we extract a yield enhancement of more than one order of magnitude at 10.9 eV and 14 eV. Extrapolating from the conversion efficiencies of 10\textsuperscript{-6}-10\textsuperscript{-7} that have been reported for silica and single-color HHG with a near-infrared driver in the same photon energy range \cite{luu18a}, this places the conversion efficiency of wave mixing from silica well into the 10\textsuperscript{-5}-10\textsuperscript{-6} range, which is comparable to gas-phase sources.
The aforementioned laser-driven carrier motion alone - which is a hallmark of mechanisms discussing both intraband currents and interband polarizations - does not account for this yield enhancement. 
For a more in-depth investigation what causes this unusual high efficiency of WMOs, we first verify experimentally the relative importance of both driving wavelengths. The separation of fundamental and SH photon channels in WMOs allows to identify that harmonic generation is more efficient for short-wavelength drivers. If we generate with both the fundamental and SH field at equal intensities (each 5.4 TW/cm\textsuperscript{2}), we observe that the dominant harmonic emission occurs within WMOs that are along the 400-nm divergence direction shown in Fig.~\ref{fig:fig4}a, signaling a dominant contribution of the 400-nm field to the WM pattern. Simulations for equal intensity of 800-nm, 400-nm or two-color pulses in Fig.~\ref{fig:fig4}b indeed confirm that aiding HHG by adding a shorter wavelength leads to the highest harmonic emission.
Comparing the momentum and time-resolved photocarrier excitation for 800-nm and 400-nm pulses in Figs.~\ref{fig:fig4}c,d, we make two observations: First, the use of a short driving wavelength increases the ionization probability and thus the number of excited carriers. Second, the shorter driving wavelength accelerates carriers across $k$ with a higher frequency, leading to enhanced carrier concentration around the $\Gamma$-point, where the band curvature facilitates both large carrier velocity and acceleration. 
This combination of increased photoexcitation, and high-frequency light-induced carrier motion results in a larger polarization for shorter driving wavelengths, giving more efficient high-order harmonic emission in the XUV, shown for 800~nm and 400~nm in Figs.~\ref{fig:fig4}e,f, respectively. 
Mixing two driving fields additionally leads to the advantageous  combination of high yield and broad spectral coverage.
The driving force of enhanced carrier concentration is most clearly seen from the $k$-resolved and energy-resolved polarizations in Figs.~\ref{fig:fig4}e,f. Driving HHG with 400 nm (Fig.~\ref{fig:fig4}f) supports Floquet-Bloch states\cite{wang13a} where the bands are dressed with multiples of the driving photon energy. Transitions between such states gives rise to the polarization signals at energies of multiples of the $k$-dependent band energies, clearly visible in Fig.~\ref{fig:fig4}f. This contrasts to the situation with 800~nm, where the strong driving and longer wavelength allows for the carriers to accelerate beyond the Brillouin zone edge. Therefore Wannier-Stark states - the eigenstates in a quasi-static electric field of an electron undergoing Bloch oscillations - are a more appropriate basis where dressing by multiples of the Bloch frequency occurs \cite{higuchi14a}. 
This illustrates that the shorter driving field operates further in the multiphoton regime providing assistance to more efficient carrier excitation and thus overall boosting HHG emission intensity. Microscopically, this has important ramifications that can stimulate new theories: While Floquet-Bloch dressing with 400~nm drivers supports a delocalized excitation due to dressing of the bands and thus encompassing many unit cells, the strong 800~nm driver leads to subsequent spatial localization due to Wannier-Stark dressing, and emission frequencies given by the energy differences of the localized Wannier-Stark states.

In conclusion, we have demonstrated efficient extreme-ultraviolet two-color non-collinear high-order wave mixing in silica.
Our theory based on semiconductor Bloch equations revealed that both Floquet-Bloch dressing followed by interband and intraband dynamics, that can rather be rationalized as dressed Wannier-Stark states, compete during intense laser-irradiation promoting high-efficiency carrier excitation and thus boosting the HHG emission intensity.
The resulting brightness enhancement turns solid HHG into a competitive source, especially in combination with the recently pioneered wavefront shaping of solid HHG through nanostructures \cite{roscamCLEO21}. Moreover, our simulations show that harmonics can be understood as signatures of laser dressing and Floquet engineering of materials \cite{perezpiskunow14a,oka19a}, a topic that has recently gathered much attention due to the opportunity to modify electronic structures of solids at will purely through light-matter interaction \cite{lindner11a,mciver20a}.
In addition, the multiband dynamics that drive wave mixing in this study correspond to a field-driven attosecond electron wave packet motion in the crystal lattice. The simultaneous recording of multiple experimental observables from solid HHG may thus allow for the reconstruction of this attosecond electron motion as previously achieved in molecules \cite{kraus15b}, and become a valuable tool for investigating laser-driven phase transitions and strongly correlated electron dynamics trough high-harmonic spectroscopy \cite{silva18a}, with the possibility to add picometer spatial resolution as demonstrated for solid HHG recently \cite{lakhotia20a}. 

\section*{Methods}
\textbf{Sample preparation.} The 100 \textmu m-thick amorphous fused silica sample utilized in this experiment was purchased from UQG Optics (CFS-1010 UV Coverslip) and cleaned with an acid piranha (H$_2$O / 37\% NH$_4$OH / 30\% H$_2$O$_2$, 5:1:1 mixture) solution at 70 degrees for 15 minutes, followed by isopropanol cleaning, prior to experimentation.
\newline

\textbf{Experimental setup.} The output of a Ti:sapphire amplifier (Coherent Astrella, 800 nm, 1 kHz) is split into two arms. In one arm, the second harmonic is generated inside a 200 \textmu m thick \textbeta-barium borate (BBO) crystal. The second arm contains a delay line to optimize temporal overlap of the two pulses onto the sample. Half-wave plates are placed in both arms to ensure parallel polarization of both pulses. Focusing is achieved by means of a spherical mirror with focal length f=500 mm. The crossing angle of 17.5 mrad in focus is implemented by impinging onto the spherical mirror with the two beams parallel and displaced in height by 9 mm.
The beam sizes are individually adjusted using irises so as to maximize the signal while preventing damaging the sample. This leads to a focal distribution of the second harmonic to be half the size of the fundamental at full-width at half maximum. 
The generated harmonics are spectrally dispersed along the horizontal direction with an aberration-corrected, concave, flat-field grating (1200 lines/mm) and detected by a double-stack micro-channel plate assembly, backed with a phosphor screen, which is imaged with a CMOS camera from outside the vacuum chamber.
\newline

\textbf{Simulations} We model HHG from fused silica as a strongly-driven system of one valence band and two conduction bands for a 40-fs pulse along the $\Gamma$-M crystal direction. The energies and the strength of the dipole coupling are calculated within density functional theory, using pseudopotentials and a plane-wave basis set as implemented in Quantum Espresso. The high harmonic emission is calculated by solving the semiconductor Bloch equations. The far field pattern is obtained by propagating the emission pattern to the far field by Fraunhofer diffraction. Details are given in the SI, section II.

\section*{Data availability}
The data that support the plots within this paper and other findings of this study
are available from P.M.K. upon reasonable request.

\section*{Code availability}
The codes that support the findings of this study are available from P.M.K. upon reasonable request.

\section*{Acknowledgements}
This work was carried out at the Advanced Research Center for Nanolithography (ARCNL), a public-private partnership of the University of Amsterdam (UvA), the Vrije Universiteit Amsterdam (VU), the Netherlands Organisation for Scientific Research (NWO), and the semiconductor equipment manufacturer ASML. Part of this work was carried out on the Dutch national e-infrastructure with the support of SURF Cooperative.
We thank Stefan Witte and Wim Ubachs for the critical review of the manuscript.
We thank Reinout Jaarsma for technical support. We thank the mechanical workshop and the design, electronic, and software departments of ARCNL for the construction of the setup. P.M.K. acknowledges support from the Netherlands Organisation for Scientific Research (NWO) under Veni Grant No. 016.Veni.192.254.

\section*{Author contributions}
S.D.C.R.A., F.C, and P.M.K. conceived the research. F.C., S.D.C.R.A., Z.Z., and M.v.d.G. performed the experiments. F.C. and S.D.C.R.A. analyzed the data. C.M and S.D.C.R.A developed the density functional theory calculations. S.D.C.R.A., B.d.K, and P.M.K. developed the theory and performed the simulations regarding high harmonic emission. P.M.K. supervised the research. S.D.C.R.A, F.C., and P.M.K. wrote the manuscript with input from all authors.

\section*{Competing interests}
The authors declare no competing interests.

\end{document}



\title{Supplementary Information: \\ Efficient extreme-ultraviolet high-order wave mixing from laser-dressed silica}
\author{Sylvianne D.C. Roscam Abbing$^{1}$}%
\email{s.roscam@arcnl.nl}

\author{Filippo Campi$^{1}$}
\author{Brian de Keijzer$^{1}$}

\author{Corentin Morice$^{2}$}

\author{Zhuang-Yan Zhang$^1$}
\author{Maarten L.S. van der Geest$^1$}
\author{Peter M. Kraus$^{1,3}$}
\email{p.kraus@arcnl.nl}
\affiliation{%
\authormark{1} Advanced Research Center for Nanolithography,\\ Science Park 106, 1098 XG Amsterdam, The Netherlands\\
\authormark{2} Institute for Theoretical Physics Amsterdam and Delta Institute for Theoretical Physics, University of Amsterdam, 1098 XH Amsterdam, Netherlands \\
\authormark{3} Department of Physics and Astronomy, and LaserLaB, Vrije Universiteit, De Boelelaan 1081, 1081 HV Amsterdam, The Netherlands \\
}

\date{\today}
\maketitle

\section{Selection rules and emission angle for high-order wave mixing}

Three selection rules must be considered to understand non-collinear wave mixing spectra driven by the fundamental and its second harmonic, and to assign each harmonic order to a unique combination of photons. 
Both wavelengths contribute to the emitted photon energies via an integer number of photons. From a simple energy conservation standpoint, the expected emission takes place at photon energies: $q\omega_f=n\omega_f+m2\omega_f=(n+2m)\omega_f$, where $q=n+2m$ is the harmonic order, $\omega_f$ is the angular frequency of the fundamental laser field, and $n$ and $m$ are the numbers of photons from the fundamental and its second harmonic, respectively. This yields a discrete spectrum, shown in Fig.~1b. Emission corresponding to even harmonics of the fundamental is present, despite the fact that fused silica is a centrosymmetric material. In fact, in a two-color field of commensurate frequencies, the parity conservation rule, which requires the total number of photons $n+m=$ to be odd, allows for generation of even harmonics of the fundamental through the addition (subtraction) of an odd amount of photons. The third conservation rule needed for a description of the full far-field profile shown in Fig.~1b, is conservation of the momentum. The harmonic spectrum spans less then an octave in energy, resulting into a unique mapping of each wave-mixing order (WMO) onto its emission angle $\beta$ for a crossing angle $\alpha$ between fundamental and the second harmonic,
\begin{equation}
\beta(n,m)=\frac{2m\sin\alpha}{n+2m\cos\alpha}.
\label{eq:beta}
\end{equation}
A small-angle approximation is used in Eq.~\ref{eq:beta}.
Similar concepts have previously been applied in gas-phase HHG \cite{bertrand11a,heyl14a}.

\section{Theory of high-order wave-mixing spectra}
\subsection{Semiconductor Bloch equations}
We model the generation of high-harmonics from silica, by solving the semiconductor Bloch equations \cite{lindberg1988a,golde2008high,schubert14a} for a three-level system, consisting of one valance band and two conduction bands. Throughout this manuscript we use atomic units. 
The system of coupled differential equations are solved for the momentum $k$ dependent population $f^\lambda_k$ in band $\lambda$ and the polarization $p^{\lambda\lambda'}_k$ between bands $\lambda$ and $\lambda'$, and is written out for bands $\lambda \in \{h_1,e_1,e_2\}$ as:

\begin{equation}
    i\frac{\partial}{\partial t}p^{h_1e_1}_k=(\epsilon^{e_1}_k+\epsilon^{h_1}_k-i\frac{1}{T_2})p^{h_1e_1}_k - (1-f^{e_1}_k-f^{h_1}_k)d^{e_1h_1}_kF(t)+ iF(t)\cdot \nabla_kp^{h_1e_1}_k + F(t)[d^{e_2h_1}_kp^{e_2e_1}_k-d^{e_1e_2}_kp^{h_1e_2}_k]
\end{equation}

\begin{equation}
    i\frac{\partial}{\partial t}p^{h_1e_2}_k=(\epsilon^{e_2}_k+\epsilon^{h_1}_k-i\frac{1}{T_2})p^{h_1e_2}_k - (1-f^{e_2}_k-f^{h_1}_k)d^{e_2h_1}_kF(t)+ iF(t)\cdot \nabla_kp^{h_1e_2}_k + F(t)[d^{e_1h_1}_kp^{e_2e_1}_k-d^{e_1e_2}_kp^{h_1e_1}_k]
\end{equation}

\begin{equation}
    i\frac{\partial}{\partial t}p^{e_1e_2}_k=(\epsilon^{e_2}_k-\epsilon^{e_1}_k-i\frac{1}{T_2})p^{e_1e_2}_k + (f^{e_2}_k-f^{e_1}_k)d^{e_2e_1}_kF(t)+ iF(t)\cdot \nabla_kp^{e_1e_2}_k + F(t)[d^{e_1h_1}_kp^{e_2h_1}_k-d^{h_1e_2}_k(p_k^{h_1e_1})^*] 
\end{equation}
\begin{equation}
   \frac{\partial}{\partial t}f^{e_1}_k= - 2~ \text{Im}[d^{e_1e_2}_kF(t)(p^{e_2e_1}_k)^* + d^{e_1h_1}_kF(t)(p^{h_1e_1}_k)^*]+F(t)\cdot \nabla_kf^{e_1}_k. 
\end{equation}
\begin{equation}
   \frac{\partial}{\partial t}f^{e_2}_k= - 2~ \text{Im}[d^{e_2e_1}_kF(t)(p^{e_1e_2}_k)^* + d^{e_2h_1}_kF(t)(p^{h_1e_2}_k)^*]+F(t)\cdot \nabla_kf^{e_2}_k. 
\end{equation}
\begin{equation}
   \frac{\partial}{\partial t}f^{h_1}_k= - 2~ \text{Im}[d^{e_1h_1}_kF(t)(p^{h_1e_1}_k)^* + d^{e_2h_1}_kF(t)(p^{h_1e_2}_k)^*]+F(t)\cdot \nabla_kf^{h_1}_k. 
\end{equation}

in which, the single particle energies of the carriers in band $\lambda$ are given by $\epsilon_k^{\lambda}$. 
The creation of polarization and population due to the presence of an electric field $F(t)$ follows from the terms that involve the transition dipole moment $d^{\lambda \lambda'}_k$. The intraband dynamics, caused by carrier acceleration by $F(t)$ within the bands, are described by the terms including $\nabla_k$.
The dephasing time of the polarization is denoted by $T_2$. The transition dipole moment $d^{\lambda \lambda'}_k$ between the bands $\lambda$ and $\lambda '$ is approximated in first order $\textbf{k}\cdot \textbf{p}$ theory \cite{haug2009quantum} as 

\begin{equation}
    d^{\lambda \lambda '}_k=d_0^{\lambda \lambda'} \frac{E_g^{\lambda 
    \lambda'}}{\epsilon_k^{\lambda}-\epsilon_k^{\lambda '}},
\end{equation}
where $d_0^{\lambda \lambda'}$ resembles the transition dipole moment between bands $\lambda$ and $\lambda'$ at the \textGamma-point, and $E_g^{\lambda \lambda'}$ the bandgap energy at the \textGamma-point between bands $\lambda$ and $\lambda'$. 
The semiconductor Bloch equations are cast as a set of coupled partial differential equations with periodic boundaries. To solve this numerical problem, the k-dimension of the equations are represented in a Fourier-series base and then solved by using spectral methods, as provided through the Dedalus project \cite{burns2020dedalus}. 

The macroscopic polarization $P(t)$ and macroscopic current $J(t)$ are calculated \cite{golde2008high} by
\begin{equation}
    P(t)=\sum_{ \lambda,\lambda',k}[d^{\lambda \lambda'}_k p^{\lambda \lambda'}_k + c.c.]
\end{equation}

\begin{equation}
    J(t)=\sum_{\lambda,k} v^\lambda(k)f^\lambda_k
\end{equation}
 with the group velocity $v^\lambda(k)$ given by
 \begin{equation}
     v^\lambda(k)=\nabla_k\epsilon^\lambda_k.
 \end{equation}

We calculate the spectral representation of the source field of interband polarization and intraband current as follows

\begin{equation}
    F_\mathrm{inter}(\omega)=\mathcal{F}[\frac{\partial}{\partial t}j_\mathrm{inter}(t)]=\mathcal{F}[\frac{\partial^2}{\partial^2 t}P(t)]=\omega^2\mathcal{F}[P(t)]=\omega^2P(\omega).
\end{equation}
and the current as
\begin{equation}
    F_\mathrm{intra}(\omega)=\mathcal{F}[\frac{\partial}{\partial t}j_\mathrm{intra}(t)]=i\omega\mathcal{F}[J(t)]=i\omega J(\omega).
\end{equation}
The harmonic spectral density at the sample plane is then defined as

\begin{equation}
    I_{\mathrm{HHG}}(\omega)=|\omega^2P(\omega)+i\omega J(\omega)|^2.
\end{equation}

\subsection{Far-field propagation}
The SBE yield the spatial distribution of the complex electric field for the interband and intraband contributions $F_\mathrm{inter}(\omega)$ and $F_\mathrm{intra}(\omega)$, respectively, at the sample plane. Detection is performed in the far-field which, in the Fraunhofer diffraction regime, is related to the sample plane through a Fourier transform:

\begin{equation}
\mathcal{F}[F_\mathrm{HHG}(x_{SP},\omega)]\propto F_\mathrm{HHG}(x_{FF},\omega).
\end{equation}

The field at the sample plane is described by a one-dimensional spatial variable $x_{SP}$. To obtain a far-field pattern, the harmonic signal $F_\mathrm{HHG}=F_\mathrm{inter}+F_\mathrm{intra}$ is computed for every value of $x_{SP}$. The near-field spectrum of the solid HHG is propagated to the far-field by Fraunhofer diffraction, which gives the wave mixing pattern shown in Fig.~2(b) in the main text.
The total field is built up by an 800-nm pulse and a 400-nm pulse, which are crossed under an angle $\beta$, thus creating a time-dependent grating-like interference pattern. The FWHM size of the focus of the 800-nm pulse is measured to be 170~\textmu m, the 400-nm focus is measured to be 70~\textmu m. 
In the focal plane both pulses are assumed to have a flat spatial phase front. The only remaining phase factor is introduced by the angle $\beta$ between the two pulses, which depends on the position x, $\phi(x) \approx x \sin(\beta)$. 

\subsection{Separation into band-resolved interband-polarization and intraband-current contributions}
\begin{figure} [h!]
\centering
	\includegraphics[width=\textwidth]{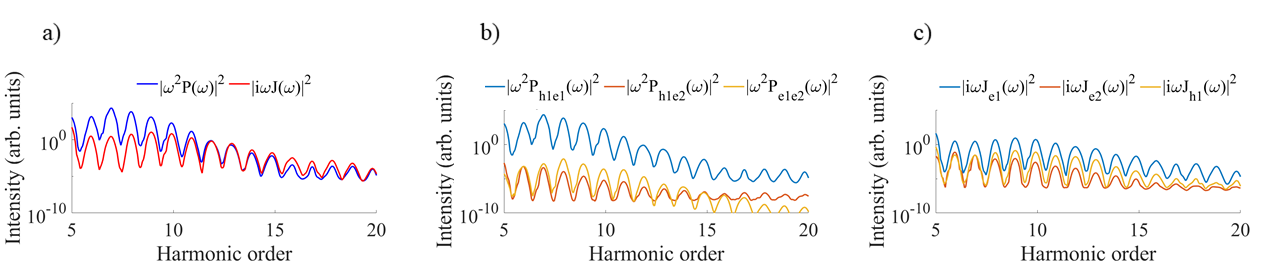} 
	\caption{a) Total interband polarization and intraband current. b) Separate contributions of the interband polarization. c) Separate contributions of the intraband current.}
	\label{fig:PJ_WM}
\end{figure}

The total harmonic signal in the wavemixing configuration of Fig.~2b (main text) is calculated by summing up all contributions along the sample plane, see Fig.~S\ref{fig:PJ_WM}a. To identify which bands contribute most to the signal, we calculate the separate values of the interband polarization (Fig.~S\ref{fig:PJ_WM}b) and the intraband current (Fig.~S\ref{fig:PJ_WM}c)). Overall, the polarization dominates the emission in the spectral region around the bandgap (H5-H11). The main contribution of the intraband current originates from the carriers inside the first conduction band. The main contribution of the polarization is generated between the valence and the lowest lying conduction band. 

\subsection{Electronic-structure calculations for material parameters of Silica }
The energy levels and the transition dipole moments of fused silica at the \textGamma-point between the valence band and two conduction bands are calculated within density functional theory, using pseudopotentials (generalised gradient approximation in the shape of the Perdew, Burke, and Ernzerhof (PBE) functional \cite{perdew1996generalized}) and a plane-wave basis set as implemented in Quantum Espresso \cite{QE-2017}. The plane-wave energy cutoff used was 40 Ry. The calculation is performed for an automatically generated uniform grid of 10 points in each k direction. Both these parameters were tested for convergence. The energy levels at $\mathrm{k=0}$ are calculated to be 0 eV, 9.6 eV and 12.6 for the valence and the two lowest lying conduction bands respectively. The dipole moments between these energy levels $\lambda,\lambda'$ are calculated to be $d_0^{h_1e_1}=0.318$ a.u., $d_0^{e_1e_2}=0.014$ a.u. and $d_0^{e_2h_1}=0.037$ a.u., using the epsilon function from the post processing data package. The dispersion relation of the bands of silica are taken from \cite{luu15a}. We model the light-matter interaction along the $\Gamma$-M crystal direction of quartz, as this orientation generates the most harmonic emission \cite{luu15a}. 

\subsection{Semiclassical description of laser-driven carrier motion}
The semiclassical description \cite{wegener2005extreme, luu15a} of the carrier dynamics inside an energy band $\lambda$ starts with describing the velocity of an electronic wavepacket by
\begin{equation}
    \textbf{v}^\lambda(\textbf{k},t)=\frac{\partial \epsilon^\lambda(\textbf{k})}{\partial \textbf{k}}.
\end{equation}
The crystal momentum is time-dependent, given by
\begin{equation}
    \textbf{K}(t)=\textbf{k}+\textbf{A}(t),
\end{equation}
with the vector potential, 
\begin{equation}
    \textbf{A}(t)=-\int_{\infty}^{t}\textbf{F}(t')dt'
\end{equation}
based on the two-color laser field $\textbf{F}(t)=F_{800}\cos(\omega t) + F_{400}\cos(2\omega t + \phi)$, with the field amplitudes of the fundamental ($F_{800}$) and second harmonic ($F_{400}$) field, and the relative phase $\phi$ between the fields.
By expressing the dispersion relation of band $\lambda$ as,

\begin{equation}
    \epsilon^{\lambda}(\textbf{k})=\sum_{\textbf{r}} E^\lambda(\textbf{r})\cos(\textbf{k}\cdot \textbf{r})
\end{equation}

in which the band coefficients $E^\lambda(\textbf{r})$ represent the amplitude of the spatial harmonics of the lattice structure, and by substituting the time-dependent crystal momentum we obtain a final expression for the velocity
\begin{equation}
    \textbf{v}^\lambda (\textbf{k},t)=-\sum_{\textbf{r}}\textbf{r}E^{\lambda}(\textbf{r})\sin[(\textbf{k}+\textbf{A}(t))\cdot\textbf{r}].
\end{equation}
We calculate the harmonic intensity as
\begin{equation}
    I_\mathrm{HHG}(\omega) \propto \sum_{\textbf{k}}|\omega \mathcal{F}[\textbf{v}^\lambda(\textbf{k},t)]|^2.
\end{equation}
To reproduce the intensity scalings, we calculate the spectrum for each combination of intensities of the fundamental and second harmonic in Fig.~3 of the main text. The semiconductor Bloch simulations reveal that the main contribution is coming from the polarization between the valence band and the first conduction band. In addition, the nonlinearity of the valence band is much lower than for the first conduction band, such that the main emission will originate from the first conduction band. Therefore we calculate the spectrum used in the intensity scalings only based on the dispersion relation of the first conduction band. 

\section{Additional experimental data and analysis}
\subsection{Intensity scaling of the harmonic yield}
\begin{figure} [h!]
\centering
	\includegraphics[scale=0.5]{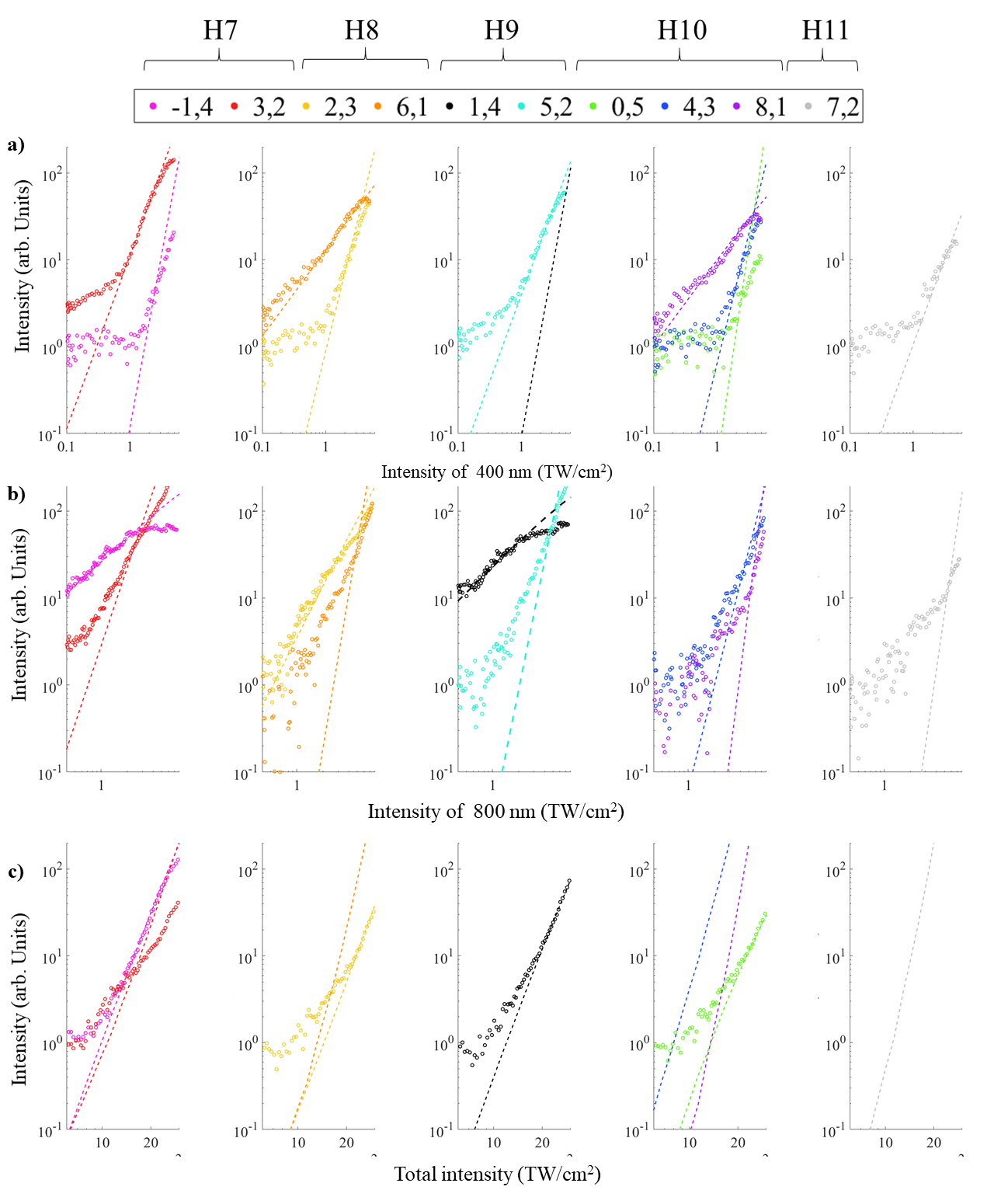} 
	\caption{Comparison between experimental data (open circles) and simulations (dashed lines). a) The intensity of the 800-nm pulse is 11.0 TW/cm\textsuperscript{2}. b) The intensity of the 400-nm pulse is 5.4 TW/cm\textsuperscript{2}. c) The ratio of the intensities of the 800-nm pulse over that of the 400-nm pulse is 1:3.5.}
	\label{fig:S2}
\end{figure}

The experimental intensity is calculated using the measured pulse length, refractive index of the material, focus size and pulse energy. The pulse length for the fundamental pulse was determined by a home-built frequency-resolved optical gating (FROG) setup and the focus sizes were measured outside the chamber, where focal overlap was maximum. 
Figure S2 shows all measured WMOs, a selection of those is shown in Fig. 3. The measured data is displayed as circles and the simulated data, based on the semiclassical description of laser-driven carrier motion (SI, section II.E), as dashed lines. Not all experimental intensity scalings show all WMOs, since the appearance of WMOs depends on the combination of fundamental and 400-nm intensity used. For the scaling of the 400-nm pulse intensity, we observe that for all WMOs there are intensity ranges which are accurately described by a perturbative power scaling. 
As mentioned in the main text, during the intensity scaling of the fundamental pulse (see Fig. S2b), the overlap between both two foci may have deteriorated for higher intensities due to a thermal drift. This leads to an overestimation of the fundamental intensity involved in the wavemixing, and therefore shows deviating power laws at higher intensities. 

\newpage

\subsection{Enhancement factors}
In all experimental scans the WMOs show the highest signal compared to the on-axis harmonic orders. In order to estimate how much brighter the WMOs are, we compare the yield of a WMO with the yield of an on-axis harmonic generated at the same photon energy and total intensity, see Fig.~S3. To obtain the fairest comparison, we compare scans where HHG is exclusively driven by the fundamental 800 nm pulses and in which the total intensity is build up from both driving wavelengths. Both scans were taken under similar experimental conditions.  

More specifically, we compare scans where the intensity of the 800-nm pulse is around 11.0 TW/cm\textsuperscript{2} and the intensity of the 400-nm pulse is varied (blue lines in Fig.~S3). At this 800~nm intensity, the efficiency of HHG driven by 800~nm alone is close to maximal and the main on-axis harmonics (orders (7,0), (9,0), and (11,0)) are always present, and their yield does not change significantly by addition of the second color.
Scans in which the intensity of the 800-nm pulses is scaled and the intensity of 400~nm is kept constant (red lines in Fig.~S3) start at lower intensity of the 800~nm pulses and the WMOs show a dramatic increase in yield. Comparing the yield of the WMOs as a function of total driving intensity (fundamental and SH intensity combined) with the yield of on-axis harmonic orders at the same photon energy and total driving intensity, provides the enhancement factors. 
The yield of order (7,0) is only a few arbitrary units at an intensity of 11.0 TW/cm\textsuperscript{2}. At the same total intensity, which now consist of both fundamental and SH intensity, the signal of WMO (3,2) has a yield of about 170 (arbitrary units), leading to an enhancement factor of 85. An enhancement factor of about 50 is found for H9, shown in Fig.~S3b, in which the on-axis harmonic (9,0) has a single-digit yield, whereas the WMO (5,2) shows a yield of around 100. For H11 the enhancement is least pronounced, the signal of the WMO (7,2) is about 7 times larger then the on-axis harmonic (11,0). Overall we are confident to report an increase of the WMOs with at least one and up to two orders of magnitude compared to the on-axis harmonics for H7 and H9, and about half an order of magnitude for H11. 
\begin{figure} [h!]
\centering
	\includegraphics[scale=0.5]{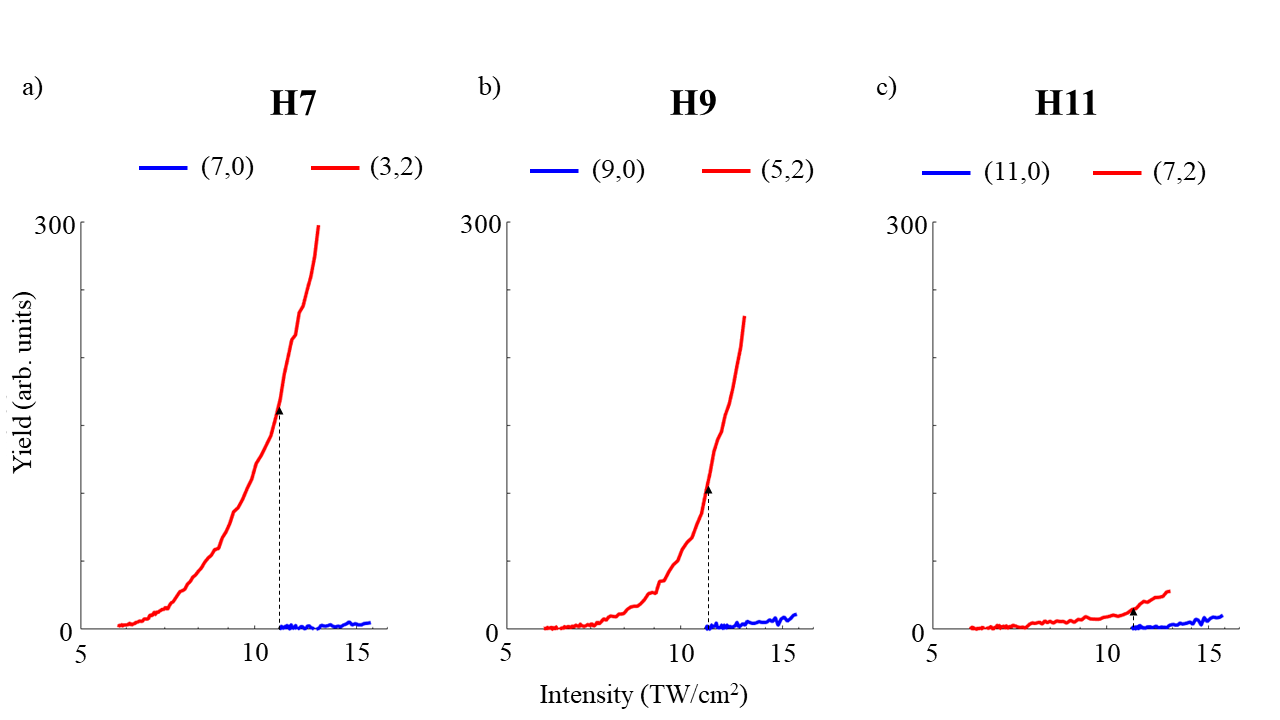} 
	\caption{The yield of several harmonic orders (of the fundamental) at a) 10.9~eV, b) 14.0~eV, and c) 17.1 eV photon energy, as a function of total driving intensity. The blue curves represent on-axis harmonics, in which the 800-nm intensity is fixed around 11.0 TW/cm\textsuperscript{2} and the 400-nm intensity is increased from 0 TW/cm\textsuperscript{2} to about 5 TW/cm\textsuperscript{2}. The red curves represent WMOs at the same photon energies (10.9 eV, 14.0 eV, and 17.1 eV, respectively) in which the 400-nm intensity is fixed around 5.4 TW/cm\textsuperscript{2} and the 800-nm intensity is increased from 0 to 7.5 TW/cm\textsuperscript{2}.}
	\label{fig:Enh}
\end{figure}